\documentclass[12pt,a4paper]{article}
\usepackage{a4wide}
\usepackage{amsfonts}
\usepackage{amssymb}
\usepackage[notref,notcite]{}

\begin{document}

\newtheorem{theo}{Theorem}
\newtheorem{lemma}{Lemma}
\newtheorem{cor}{Corollary}
\newtheorem{defi}{Definition}
\newtheorem{prop}{Proposition}

\newcommand{\proofend}{\raisebox{1.3mm}{%
\fbox{\begin{minipage}[b][0cm][b]{0cm}\end{minipage}}}}
\newenvironment{proof}[1][\hspace{-1mm}]{{\noindent\it Proof #1:}
}{\mbox{}\hfill \proofend\\\mbox{}}

\def\inser{\lrcorner}
\def\be{\begin{equation}}
\def\ee{\end{equation}}
\def\re{(\ref }
\def\tr{{\rm tr \, }}

\let\a=\alpha \let\b=\beta \let\g=\gamma \let\d=\delta
\let\e=\varepsilon \let\ep=\epsilon \let\z=\zeta \let\h=\eta \let\th=\theta
\let\dh=\vartheta \let\k=\kappa \let\l=\lambda \let\m=\mu
\let\n=\nu \let\x=\xi \let\p=\pi \let\r=\rho \let\s=\sigma
\let\t=\tau \let\o=\omega \let\c=\chi \let\ps=\psi
\let\ph=\varphi \let\Ph=\phi \let\PH=\Phi \let\Ps=\Psi
\let\O=\Omega \let\S=\Sigma \let\P=\Pi
\let\Th=\Theta \let\L=\Lambda \let \G=\Gamma \let\D=\Delta

\def\({\left(} \def\){\right)} \def\<{\langle} \def\>{\rangle}
\def\lb{\left\{} \def\rb{\right\}}
\def\[{\lbrack} \def\]{\rbrack}
\let\lra=\leftrightarrow \let\LRA=\Leftrightarrow
\def\ul{\underline}
\def\wt{\widetilde}
\let\Ra=\Rightarrow \let\ra=\rightarrow
\let\la=\leftarrow \let\La=\Leftarrow

\def\CG{{\cal G}}\def\CN{{\cal N}}\def\CC{{\cal C}}
\def\CL{{\cal L}} \def\CX{{\cal X}} \def\CA{{\cal A}} \def\CE{{\cal E}}
\def\CF{{\cal F}} \def\CD{{\cal D}} \def\rd{\rm d}
\def\rD{\rm D} \def\CH{{\cal H}} \def\CT{{\cal T}} \def\CM{{\cal M}}
\def\CI{{\cal I}}
\def\CP{{\cal P}} \def\CS{{\cal S}} \def\C{{\cal C}}
\def\CR{{\cal R}}
\def\CO{{\cal O}}

\newcommand{\md}{\mathrm{d}}

\newcommand*{\dId}{{\mathchoice%
{\mathrm{1\mkern-4.3mu I}}%
  {\mathrm{1\mkern-4.3mu I}}%
{\mathrm{1\mkern-3.1mu I}}%
  {\mathrm{1\mkern-3.1mu I}}}}
\newcommand*{\dR}{{\mathbb R}}
\newcommand*{\dN}{{\mathbb N}}
\newcommand*{\dZ}{{\mathbb Z}}
\newcommand*{\dC}{{\mathbb C}}
\newcommand{\Diff}{\mbox{\rm Diff}}
\newcommand{\diff}{\mbox{\rm diff}}
\newcommand{\semidir}{\mathrm{%
\times\mkern-3.3mu\protect\rule[0.04ex]{0.04em}{1.05ex}\mkern3.3mu\mbox{}}}
\newcommand{\dd}[1]{\frac{d}{d#1}}
\newcommand{\pd}[1]{\frac{\partial}{\partial #1}}
\newcommand{\zq}{\overline{z}}
\def\tl{{\widetilde \lambda}}

\newcommand{\TS}[1]{{\bf\boldmath \framebox{Th} :#1}}
\newcommand{\An}[1]{{\bf\boldmath \framebox{An}: #1}}

\def\2{\frac12}
\def\pl{\plabel}
\def\Pf{{\it Pf\/}}
\def\Ad{{\it Ad\/}}
\def\Lieg{{\it Lie\/}G}
\def\tr{{\it tr\/}} 
\def\ad{{\it ad\/}}
\def\ba{\begin{eqnarray}}
\def\ea{\end{eqnarray}}
\def\bbbc{{\mathchoice {\setbox0=\hbox{$\displaystyle\rm C$}\hbox{\hbox
to0pt{\kern0.4\wd0\vrule height0.9\ht0\hss}\box0}}

{\setbox0=\hbox{$\textstyle\rm C$}\hbox{\hbox
to0pt{\kern0.4\wd0\vrule height0.9\ht0\hss}\box0}}
{\setbox0=\hbox{$\scriptstyle\rm C$}\hbox{\hbox
to0pt{\kern0.4\wd0\vrule height0.9\ht0\hss}\box0}}
{\setbox0=\hbox{$\scriptscriptstyle\rm C$}\hbox{\hbox
to0pt{\kern0.4\wd0\vrule height0.9\ht0\hss}\box0}}}}
\def\empty{{\emptyset}}

 \hfill  FSU-TPI-07/04
{\renewcommand{\thefootnote}{\fnsymbol{footnote}}
\medskip
\begin{center}
{\LARGE \bf Current Algebras and Differential Geometry}

\vspace{1.5em}
Anton Alekseev \\ 
Section of Mathematics, University of Geneva, \\
2-4 rue du Lievre, c.p. 240, 1211 Geneva 24, Switzerland;\\
Institute for Theoretical Physics, Uppsala University, \\
Box 803, 75108 Uppsala, Sweden
\\[3.5mm]
Thomas Strobl\\ 
Institut f\"ur Theoretische Physik, Universit\"at Jena\\
D--07743 Jena, Germany\\
\vspace{1.5em}

\emph{We dedicate this article to Ludwig Faddeev on
the occasion of his 70th
birthday.}

\vspace{0.5em}

\begin{abstract}
We show that symmetries and gauge symmetries of a large class of
2-dimensional $\sigma$-models are described by a new type of
a current algebra. The currents are labeled by pairs of a 
vector field and a 1-form on the target space of the $\sigma$-model.
We compute the current-current commutator and analyse the
anomaly cancellation condition, which can be interpreted 
geometrically in terms of {\em Dirac structures},
previously studied in the mathematical literature. 
{\em Generalized complex structures} correspond to 
decompositions of the current algebra into pairs of anomaly
free subalgebras. $\sigma$-models that we can treat with our method
include both  physical and topological examples, with
and without Wess-Zumino type terms. 
\end{abstract}
\end{center}}
\setcounter{footnote}{0}

\vspace{-5mm}

\tableofcontents

\section{Introduction}

Current algebras originate in particle physics \cite{Treiman:1986ep}. 
The minimal coupling in gauge theories has
the form $A^a_\mu J_a^\mu$, where $A_\mu^a$ is the
gauge field and  $J^\mu_a = Tr \, e_a (\bar{\psi}
\gamma^\mu \psi)$ is the fermionic current.
Here $\psi$ and $\bar{\psi}$ are
fermionic fields, $\gamma^\mu$ are $\gamma$-matrices
and $e_a$ are the Lie algebra generators,
$[e_a, e_b]=-f_{ab}^c e_c$. 
The understanding of properties of the currents $J^\mu_a$ is an
essential piece in understanding the gauge
coupling. In particular, by computing the density-density
commutators $[J^0_a(x), J^0_b(y)]$ one usually finds
an expression of the follwoing type,
\be \label{eq:basic}
[J^0_a(x), J^0_b(y)] = f_{ab}^c J^0_c(x) \delta(x-y) 
+ {\rm anomalous \, terms}
\ee
where anomalous terms contain derivatives of $\delta(x-y)$.
In gauge theory, anomalous terms indicate that the gauge
symmetry cannot be preseved at the quantum level. If the anomalous
terms are absent, the currents $J^0_a$ form a set of first class constraints
which can be imposed on the quantum system 
\cite{Faddeev:1984jp, Faddeev:1985iz}. 

Non-abelian one-dimensional current algebras
or Kac-Moody algebras,
\be \label{eq:KacMoody}
[J_a(x), J_b(y)] = f_{ab}^c J_c(x) \delta(x-y) 
+ \kappa \delta_{ab} \delta'(x-y) .
\ee
can be viewed as coming from the theory of (1+1)-dimensional 
fermions interacting with a non-abelian
gauge field. The algebra (\ref{eq:KacMoody}) plays 
a crucial role in Conformal Field Theory
as the symmetry of the Wess-Zumino-Witten model \cite{Knizhnik:1984nr}.
It is also subject of a well developed
mathematical theory \cite{Kac}. 

In this paper we shall construct new current algebras
of the type (\ref{eq:KacMoody}) with an index $a$
replaced by a pair formed by a vector field and a 1-form on a 
manifold $M$ which serves as a target of a $1+1$-dimensional
$\sigma$-model. These current algebras naturally arise
in the description of symmetries and gauge symmetries
of both topological and physical $\sigma$-models.
In fact, several examples of our current algebra,
including (\ref{eq:KacMoody}), were known before,
but we now present a unifying picture for many 
different types of $\sigma$-models. As an interesting twist in 
our calculation, we find a relation between the new 
current algebras and \emph{Courant brackets}
studied in the mathematical literature \cite{WeinsteinCourant}. 
Some relation between Courant brackets and current (or vertex)
algebras has been previously considered in \cite{MS},
\cite{Bres}.

In Section 2 we recall the Lagrangian and Hamiltonian
description of several classes of 2-dimensional $\sigma$-models.
Our list includes the WZW model, the gauged WZW model, 
Poisson and WZ-Poisson $\sigma$-models. In Section 3
we compute the current-current commutator and express
it in terms of the Courant bracket. While the commutator 
in the full current algebra always has an anomalous part,
for subalgebras one can study the anomaly concellation
condition. We find that it gives rise to {\em Dirac structures}.
Two transversal Dirac structures form a 
{\em generalized complex structure} and give rise
to a polarization of our current algebra into two anomaly
free subalgebras.

\section{2-dimesional $\sigma$-models}

\subsection{Examples} 
In this paper we consider several classes of 
2-dimensional $\sigma$-models. Some of them
are of importance in string theory applications, and 
others are topological field theories of interest
in mathematical physics and the theory of quantization.

Let $\Sigma$ be the two-dimesnional world-sheet or space-time
(either Eucledian or Lorentzian), and $M$ the target
manifold of the $\sigma$-model. On local charts, $\Sigma$
has coordinates $x^\alpha, \alpha=1,2$ and $M$ has
coordinates $X^i, i = 1, \dots, {\rm dim} M$. 
A typical example of a $\sigma$-model is defined by 
metrics $h_{\alpha \beta}$ and $G_{ij}$ on the world-sheet and on the target
space, respectively,
\begin{equation} \label{eq:S1}
S = \int_\Sigma \frac{1}{2} G_{ij}(X) \, \md X^i \wedge * \md X^j \, 
,  
\end{equation}
where $* \a$ is the Hodge dual of $\a$ with respect to 
$h$ (and thus $ \md X^i \wedge * \md X^j = \partial_\alpha X^i 
\partial^\alpha X^i \md \mathrm{vol}_\Sigma$, with 
$\partial^\alpha X^i \equiv h^{\alpha \beta} \partial_\beta X^i$ and 
$\md \mathrm{vol}_\Sigma \equiv \sqrt{|{\mathrm{det}} {h}|}\, \md^2x$). 
Such $\sigma$-models arise in the theory of bosonic strings
(or as bosonic parts of super-string actions) as well
as in the theory of integrable models ({\em e.g.} the
2-dimensional $O(3)$ $\sigma$-model).

Given a 2-form $B$ on the target space $M$ one can
complement the action (\ref{eq:S1}) as follows,
\be \label{eq:S2}
S[X] = \int_\Sigma  \frac{1}{2}G_{ij}(X) \, \md X^i \wedge * \md X^j 
+ \int_\Sigma  \frac{1}{2} B_{ij}(X) \, \md X^i \wedge \md X^j .
\ee
Here the second term is an integral over the world-sheet
of the pull-back $X^*B$ with respect to the map
$X\colon \S \to M$. More generally, given a closed 3-form
$H$ on $M$ one can add a Wess-Zumino term to the action
(\ref{eq:S1}),
\be \label{eq:S3}
S[X] = \int_\Sigma \frac{1}{2} G_{ij}(X) \, \md X^i \wedge * \md X^j 
+ \int_N H ,
\ee
where $N$ is a 3-dimensional submanifold of $M$ with
$\partial N = X(\Sigma)$. In the string theory context,
$B$ is the NS-NS 2-form, and $H$ is the corresponding
field strength.

An interesting example of (\ref{eq:S3}) is the 
Wess-Zumino-Witten (WZW) model \cite{WZW}, 
where $M=G$ is a 
Lie group with an invariant metric (for instance,
$G$ semi-simple with metric given by the Killing
form) and $H$ the Cartan 3-form. More explicitly,
the action is given by
\be \label{eq:WZW}
S[g]= \frac{k}{8\pi} \int_\Sigma \mathrm{Tr} \, 
(g^{-1} \md g \wedge * g^{-1} \md g)  +
\frac{k}{12\pi} \int_N \mathrm{Tr} 
(g^{-1} \md g)^3 .
\ee
Here $g: \Sigma \to G$ and we denote the metric
on the Lie algebra $\mathfrak{g}$ by $\mathrm{Tr}$.

There is another class of $\sigma$-models which arises
when instead of a metric $G$ on $M$ we are given a
bi-vector $\Pi=\frac{1}{2}\Pi^{ij}(X)\partial_i 
\wedge \partial_j$. Then, the action is given by
\be \label{eq:PSM}
S[A,X]= \int_\Sigma ( A_i \wedge \md X^i + \frac{1}{2}
\Pi^{ij}(X) A_i \wedge A_j ),
\ee
where $A_i=A_{i \a}(x) \md x^\a$ and $A_{i \a}$ are 
components of 1-forms both on the world-sheet and on
the target space. This model is topological if $\Pi$
is a Poisson bi-vector, i.e.~$\Pi^{ij} := \{X^i,X^j\}$ are Poisson
brackets on $M$. Then, it is called a
Poisson $\sigma$-model \cite{proce, PSM1, Ikeda}. 

Similar to $\sigma$-models defined by a metric,
one can add a Wess-Zumino term to the action
(\ref{eq:PSM}),
\be \label{eq:HPSM}
S[A,X]= \int_\Sigma  A_i \wedge \md X^i + \frac{1}{2}
\Pi^{ij}(X) A_i \wedge A_j + \int_N H .
\ee
This action defines a topological field theory
(the space of classical solutions modulo gauge
symmetries is finite dimensional), iff $(\Pi, H)$
defines a WZ--Poisson structure, i.e.~if  
\be \label{WZPoisson}
\Pi^{il} \partial_l \Pi^{jk} + {\rm cycl}(ijk)
= \Pi^{ii'} \Pi^{jj'} \Pi^{kk'} H_{i'j'k'}
\ee
holds true. (In terms of the bivector $\Pi$ this may be  rewritten
as $ \frac{1}{2} [\Pi, \Pi] =
\langle H,\Pi \otimes \Pi \otimes \Pi \rangle$).
Then the action (\ref{eq:HPSM}) defines the 
WZ-Poisson $\sigma$-model \cite{Ctirad} (cf. also \cite{Park}).  

The WZW model can be turned 
into a topological theory \cite{Gawedzki:1988nj} too, the so-called
$G/G$ model,
by adding to the action (\ref{eq:WZW}) an extra
piece reminiscent of (\ref{eq:PSM}),
\be \label{eq:G/G}
\Delta S[g, a] = \frac{k}{4\pi}
\int_\Sigma
\mathrm{Tr} \left( a \wedge (* - 1) \md g  g^{-1} - a \wedge
(* + 1) g^{-1} \md g  - a \wedge (* - 1) ga  g^{-1} + a \wedge * a
\right)  ,
\ee
where $a$ is a  $\mathfrak{g}$-valued connection 1-form and, as before,
$*$ denotes the Hodge duality operator with respect to the world-sheet
metric $h$.

\subsection{Hamiltonian formulation}
All the models listed above share the following
phase space description. For $\Sigma = S^1 \times \dR$
the phase space is the cotangent bundle $T^*LM$ 
of the loop space $LM$. Using local coordinates 
$X^i(\s)$ and their canonical conjugates $p_i(\sigma)$,
the canonical symplectic form of the cotangent
bundle is given by
\be \label{eq:om1}
\omega = \oint_{S^1} \! \delta X^i(\s) \wedge \delta
p_{i}(\s)\, \md\s ,
\ee
where $\delta$ denotes the de Rham differential on 
the phase space $T^*LM$.

Twisting (\ref{eq:om1}) by a closed 3-form $H$ on 
$M$,  gives
\be \label{eq:om2}
\omega = \oint_{S^1} \! \delta X^i(\s) \wedge \delta
p_{i}(\s)\, \md\s + \2  \oint_{S^1} \!
H_{ijk}(X(\s)) \, \partial X^i
(\s) \, \delta X^j(\s) \wedge 
\delta X^k(\s) \, \md\s \; , 
\ee
where $\partial$ is the derivative with respect 
to $\sigma$.

{}From (\ref{eq:om2}) we read off the
Poisson brackets
\ba \{ X^i(\s) , X^j(\s') \} = 0
&,& \{ X^i(\s) , p_j(\s') \} = \delta^i_j \delta(\s -
\s')\label{brackets1}  \\ \{ p_i(\s) ,
p_j(\s') \} \label{brackets2} &=& -
H_{ijk} \partial X^k \delta(\s - \s') \, .
\ea 

In the case of ({\ref{eq:S1}) this phase space is 
complemented by the specification of a Hamiltonian
\be {\cal H} = 
\frac{1}{2}\oint_{S^1} \! \left( G^{ij}(X) p_i p_j +
G_{ij}(X)  \partial X^i \partial X^j \right) \, 
\md \s\, , \label{Ham}
\ee
where $G^{ij}$ denotes the inverse to $G_{ij}$. 
If the metric $G$ admits Killing vector fields $v_a$,
the action functional ({\ref{eq:S1}), or likewise the symplectic form
(\ref{eq:om1}) and the Hamiltonian (\ref{Ham}), 
have a symmetry
generated by the Noether currents
\be \label{eq:v}
J_a(\sigma)  = (v_a)^i(X(\sigma)) \: p_i(\s) .
\ee

In the case of the WZ-Poisson $\sigma$-model,
eq.~(\ref{eq:HPSM}),  the components of $A_i$ along the (``spatial'') 
circle $S^1$ become the momenta $p_i$, and the Hamiltonian takes the form 
\be 
{\cal H} = \oint_{S^1} \! \l_i \left( 
\partial X^i + \Pi^{ij}(X) p_j \right)\, \md\s  \,  .
\ee
Here $\l_i(\s)$ are (undetermined) Lagrange 
multipliers, the ``time'' componentes of $A_i$. 
Such a Hamiltonian enforces that the currents 
\be \label{eq:const}
J^i  = \partial X^i + \Pi^{ij} p_j  
\ee
on $T^*LM$ vanish.They are the constraints of the Hamiltonian system 
corresponding to (\ref{eq:HPSM}).

The functionals (\ref{eq:v}) and (\ref{eq:const}) are two particular
examples of the following type. Choose a vector field $v=v^i(X) \partial_i$
and a 1-form $\alpha= \alpha_i(X) \md X^i$ on $M$, and associate to them
a current,
\be \label{eq:J}
J_{(v, \alpha)}(\s) = v^i(X(\s)) p_i(\s) + \alpha_i(X(\s)) \partial X^i .
\ee
Likewise, consider a WZ-type $\sigma$-model as 
in (\ref{eq:S3}), leading to the  
symplectic form (\ref{eq:om2}). Assuming that $v$ is a Killing vector field
for the metric $G_{ij}$ which preserves the 3-form $H$, 
the Noether current need not exist. There
is an extra condition which requires that the contraction of
 $v$ with $H$ is not only closed 
but exact, i.e.~that there exists some
 1-form $\alpha= \alpha_i \md X^i$ on $M$ such that 
\be \label{eq:alpha}  
v^i  H_{ijk} =
\partial_j  \alpha_k - \partial_k  \alpha_j \, .
\ee
If this condition is satisfied, the Noether current is precisely
$J_{(v, \alpha)}$.\footnote{Associating a pair $(v,\alpha)$ (rather than only a
vector field $v$) to a symmetry of a 2-dimensional $\sigma$-model
is one of the messages of Letter 1 in \cite{Severalett}.}
There is an ambiguity in choosing 1-forms $\alpha$ solving
equation (\ref{eq:alpha}), but this is all the ambiguity for the 
Noether current corresponding to  $v$. 
This situation generalizes in a straightforward way to 
the presence of several Killing vector fields leaving (\ref{eq:S3}) invariant.
In particular, in the WZW model (\ref{eq:WZW}) the left and right
chiral currents are of the form,
%
\be \label{eq:KM}
J^L= p - \frac{k}{4\pi} g^{-1} \partial g \; , \quad  
J^R= g pg^{-1} + \frac{k}{4\pi} \partial g g^{-1}.
\ee
Here $p=p(\sigma)$ is a left-invariant momentum (a Lie algebra valued
matrix of momenta). 

Investigating the Poisson brackets and commutation relations
of $J_{(v, \alpha)}$
will be one of the main goals of this paper.

\section{Current algebra}

\subsection{Current algebra and Courant bracket}

In this Section we present the computation of the Poisson
bracket $\{ J_{(u, \alpha)}, J_{(v, \beta)} \}$ between currents, 
with $u,v$ two vector fields and $\alpha, \beta$ two 1-forms.
Presenting the answer requires the following two structures.

First, we need a symmetric scalar product on the space of vector fields
and 1-forms,
\be \label{eq:scal}
\langle (u,\alpha), (v, \beta) \rangle_+ =\alpha(v) + \beta(u).
\ee
This scalar product associates to two pairs $(u,\alpha)$ and $(v, \beta)$
a function of $X^i$. Note that the right hand side of (\ref{eq:scal})
can be both positive or negative.

The second structure is known as a {\em Courant bracket} \cite{CourantPhD, YKS, Severalett} 
and it associates to the two pairs $(u,\alpha)$ and 
$(v, \beta)$ another pair of the same type,
\be \label{eq:Courant}
[ (u,\alpha), (v, \beta) ] = ([u,v]_{\rm Lie}, L_u \beta - L_v \alpha + 
\md \left(\alpha(v)\right) 
+ H(u, v, \cdot) ).
\ee
Here $[u,v]_{\rm Lie}$ is the Lie bracket of the vector fields $u$ and $v$, 
$L_u, L_v$ stand for Lie derivatives with respect to $u$ and $v$, 
respectively,   and
$H(u,v, \cdot)$ is a 1-form obtained by contracting $H$ with $u$ and $v$.
The bracket (\ref{eq:Courant}) is not skew-symmetric. It has many interesting
properties, the most interesting one being the Leibniz identity,
\be
[ (u, \alpha), [ (v, \beta), (w, \gamma) ]] =
 [[ (u, \alpha), (v,\beta) ] , (w, \gamma) ] +
[ (v, \beta), [ (u, \alpha), (w, \gamma) ]] . \label{eq:Leibniz}
\ee
This equation is a counterpart of the Jacobi identity for non skew-symmetric
brackets.

We are now ready to present the formula for a Poisson bracket of two currents,
\be  \label{eq:main}
\{ J_{(u, \alpha)}(\s), J_{(v, \beta)}(\tau) \}=
-J_{  [ (u,\alpha), (v, \beta) ] } (\s) \delta(\s - \tau)
+ \langle(u,\alpha), (v, \beta)\rangle_+(X(\tau)) \, \delta'(\s - \tau) .
\ee
This expression shows that the currents $J_{(u, \alpha)}$ form a current
algebra, with the anomalous contribution governed by the
scalar product, and with the linear in $J$ contribution given by
the Courant bracket. 

For completeness we also compute Poisson brackets between 
currents and functions on the target space,
\be \label{eq:Jf}
\{ f(X(\tau)), J_{(u, \alpha)}(\s) \} = u(f)(X(\tau)) \delta(s-\tau) .
\ee
This equation together with the Leibniz identity for the Courant
bracket ensures the Jacobi identity of the bracket (\ref{eq:main}).
Note that the currents $J_{(u, \alpha)}(\s)$ and $f(X(\tau))$ are not
independent: $\partial f = J_{(0, \md f)}$. Using test functions
$\epsilon(\s)$, this linear dependence may be expressed also as
\be \oint  \left[\epsilon(\s) J_{(0, \md f)}(\sigma) + (\partial
\epsilon)(\s) f(X(\s))  \, \md \s \right] \equiv 0 \, . \label{eq:lin}
\ee

It should be mentioned that the bracket (\ref{eq:main}) can be
presented in many different ways by changing the argument in the anomalous
term, {\em e.g.}
\begin{eqnarray}
\{ J_{(u, \alpha)}(\s), J_{(v, \beta)}(\tau) \}&=&
-
J_{
[ (u,\alpha), (v, \beta) ] - \md \langle (u,\alpha),
(v, \beta) \rangle_+ }
(\s) \delta(\s - \tau) \nonumber \\ &&
+ \langle (u,\alpha), (v, \beta)\rangle_+ (X(\s)) \delta'(\s - \tau)
\label{eq:new} 
\end{eqnarray}    
or 
\begin{eqnarray}
\{ J_{(u, \alpha)}(\s), J_{(v, \beta)}(\tau) \}&=&-
J_{  [ (u,\alpha), (v, \beta) ] - \frac{1}{2} \md \langle
(u,\alpha), (v, \beta)\rangle_+ } (\s) \delta(\s - \tau)\nonumber \\  &&
+ \langle(u,\alpha), (v, \beta)\rangle_+\left(X( {\scriptstyle
\frac{ 1}{ 2}}(\s+
\tau))\right) \delta'(\s - \tau) . \label{eq:skew}
\end{eqnarray}
Note that the bracket that now appears in the argument of the currents,
$$
[ (u,\alpha), (v, \beta) ]_{\rm skew} =
[ (u,\alpha), (v, \beta) ] - \frac{1}{2} \md \langle (u,\alpha), (v,
\beta)\rangle_+ 
$$ 
is skew-symmetric. This is the consequence of  antisymmetry
of the Poisson brackets (\ref{brackets1}), (\ref{brackets2})
underlying the current algebra. With this bracket, however,
the nice property (\ref{eq:Leibniz}) is replaced by a 
homotopy Jacobi identity with the
right hand side given by an exact form (for details see
\cite{WeinsteinCourant}).

One can treat the linear relation (\ref{eq:lin}) in a slightly 
different way by adding the two types of currents,
$J_{(v,\a)}(\sigma)$ and $f(X(\s)))$, corresponding to an abelian
extension of the $J$-current algebra. The extended currents now form a
Lie algebra, obtained from (\ref{eq:main}) and
(\ref{eq:Jf}) above. Smearing them
by means of test functions, and using  $C^\infty(M) \otimes
C^\infty(S^1) \cong
C^\infty(M\times S^1)$, one is then lead to consider
\be  J_\psi = \oint \left( v^i(X(\s),\s) p_i(\s) + \a_i(X(\s),\s)
\partial X^i + f(X(\s),\s) \right) \md \s \,  \label{eq:ext}
\ee
as extended currents.
Here $\psi$ may be interpreted as a section of $\widetilde E :=
T\widetilde M \oplus 
T^*\widetilde M$, $\widetilde M \equiv M \times S^1$, 
whose tanget vector part is parallel to $M$: $\psi =
v^i(X,\s) \partial_i + \a_i (X,\s) \md X^i + f(X,\s) \md \s$. 
The kernel of the map $\psi \mapsto J_\psi \in C^\infty(T^*LM)$ is
provided by the exact 1-forms on $\widetilde M$, $J_{\widetilde \md f}
\equiv 0$, where $\widetilde \md = \md + \md \s \wedge \partial$ is
the de Rham differential on $\widetilde M$; this just reexpresses 
(\ref{eq:lin}).
The Lie algebra of the extended currents (\ref{eq:ext})
may now be cast into the
following simple form:
$$ \{ J_{\psi_1}, J_{\psi_2}\} = - J_{[\psi_1,\psi_2]} \, , $$
where $[\psi_1,\psi_2]$ denotes the Courant bracket in $\widetilde
E$. Indeed, modulo exact terms the
Courant bracket becomes antisymmetric, cf.~Eq.~(\ref{eq:Courant}),
so that the quotient algebra is a Lie algebra on behalf of
(\ref{eq:Leibniz}).\footnote{This quotient Lie algebra is certainly not
$C^\infty$-linear, so it cannot arise from a Lie algebroid.}
Thus the map $J_\cdot$ is an (anti-)isomorphism from the Lie algebra
constructed from the Courant bracket on $\widetilde E$ as described
above to our ``current'' Lie algebra, realized as Poisson subalgebra in
loop phase space $T^*LM$.%
\footnote{We are grateful to the Referee for suggesting this
perspective.} 

The current-current brackets (\ref{eq:main}) (or (\ref{eq:new}), 
(\ref{eq:skew}))
resemble anomalous commutators in (3+1) dimensions
\cite{Jackiwcurrents,Faddeev:1984jp, Faddeev:1985iz}:
$$
[J^0_a(x) , J^0_b(y)]= f_{ab}^c J^0_c(x) \delta(x-y) +
d_{abc} \epsilon_{ijk} \partial_i A_j^c(x) \partial_k \delta(x-y).
$$
Here $A_j^a$ is the background Yang-Mills field,
and $d_{abc}$ are symmetric structure constants
$d_{abc}=1/2 \, {\rm Tr}\, (e_ae_b+e_be_a)e_c$. 
Similar to (\ref{eq:main}),  the coefficient
in front of the derivartive of the $\delta$-function
is a field with a nontrivial $x$-dependence.

The only piece of data that we used to define the current
algebra was a closed 3-form $H$ on $M$. In fact, the current algebra
only depends on the cohomology class of $H$. Indeed, 
for two choices of a 3-form, $H$ and $H'=H+dB$, the Poisson brackets 
(\ref{brackets1}), (\ref{brackets2}) are related to one another
by a simple (non-canonical) transformation
\begin{equation} \label{eq:change}
X^i(\sigma) \mapsto X^i(\sigma) \, , \, 
p_i(\sigma) \mapsto p_i(\sigma) + B_{ij}(X(\sigma)) \partial X^j.
\end{equation}
The corresponding transformation on currents reads
$$
J_{(v, \alpha)} \mapsto J_{(v, \alpha+B(v, \cdot))} .
$$
This effect has a counterpart in mechanics of a charged
particle in a magnetic field. The transformation (\ref{eq:change}) 
is analogous to
the passage from canonical momenta $p$ to kinetic momenta $\pi= p -
\frac{e}{c} A$ for an ordinary particle in a magnetic field $B=\md A$.
Implementing this change of variables
in the canonical symplectic form $\omega = \md q^i \wedge \md p_i$, one
obtains
\be \label{eq:magnetic}
\omega  = \md q^i \wedge \md \pi_i - \frac{e}{c} B \, ,
\ee
which resembles the symplectic form (\ref{eq:om2}) on loop space
we started with. Similarly, by a shift of variables as in (\ref{eq:change}) 
we can eliminate any exact $H$ in (\ref{eq:om2}) altogether
(cf.~also \cite{Ctirad}). 

%

Finally we remark that since all $J$'s are at most linear in the
momenta $p_i$ one can consistently 
replace Poisson brackets by commutators
in all formulas above.

\subsection{Dirac structures and examples in physics}

The current algebra (\ref{eq:main}) is very big since it allows
for a choice of arbitrary vector fields and 1-forms. So, it makes sense
to look for some interesting subalgebras which are somewhat smaller.
In particular, one can ask when $J$'s form a Lie algebra with 
no anomaly term. This requires two conditions: first, all pairs
$(u, \alpha)$ in such a subalgebra should have vanishing 
scalar products, $\langle(u_1, \alpha_1), (u_2, \beta_2)\rangle=0$. Second,
the Courant brackets  should close on the space of such pairs.
If in addition $(u, \alpha)$'s  span a dimension $n=\dim M$ subbundle 
of $TM\oplus T^*M$, this is called a Dirac structure on M
\cite{CourantPhD, WeinsteinCourant}.

As the  first example, let us return to the WZ-type model
(\ref{eq:S3}) with a Killing vector field $v$ satisfying
equation (\ref{eq:alpha}) for some 1-form $\alpha$. Then,
the Noether current $J_{(v, \alpha)}(\sigma)$ has an anomalous
Poisson bracket,
$$
\{ J_{(v, \alpha)}(\sigma), J_{(v, \alpha)}(\tau) \}
= J_{(0, \md (v^i \alpha_i))}(\sigma) \delta(\sigma - \tau)
+ (v^i \alpha_i)(X(\tau)) \delta'(\sigma - \tau) .
$$
Vanishing of the anomaly gives a new condition
\be \label{eq:valpha}
 \alpha(v) \equiv v^i \alpha_i =0 \, .
\ee
Together with condition (\ref{eq:alpha}) this is tantamount 
to the 3-form $H$ extending to an equivariant 3-form,
$(\md - \iota_v)(H + \alpha)=0$, where $\iota_v$ is the contraction
with respect to the vector field $v$ (for a definition of equivariant
forms see \cite{GS}). In a similar fashion, if there are several
Killing vector fields, forming a Lie algebra
$[v_a, v_b]=- f_{ab}^c v_c$, the absence of the
$\delta'$--contribution requires $\iota(v_a)\alpha_b+
\iota(v_b) \alpha_a=0$ for all $a$ and $b$. In this case, closure of
the bracket is not automatic: in addition one needs a choice
of $\alpha$'s such that  $L_{v_a} \a_b = -f^c_{ab} \a_c$.
Then, the currents
$J_a=J_{(v_a, \alpha_a)}$ form a Lie algebra
\be
\{ J_a(\sigma), J_b(\tau) \} = f_{ab}^c J_c(\sigma) \delta(\sigma-\tau) .
\ee
Again, these conditions may be summarized compactly as saying that $H$
should extend to an equivariantly closed 3-form.

Note that even if there is an anomaly in the current-current Poisson bracket
(or commutator), the Noether charges
$Q_a = \oint \! J_a(\sigma)\, \md \sigma$  always form 
a representation of the Lie algebra; 
the anomalous term---such as any exact piece in the 1-form part of the
Courant bracket---cancels by integrating over the circle.

Anomaly free Noether currents
are needed if one wants to gauge a given rigid  symmetry. The anomaly 
cancellation condition is equivalent to requiring  that 
the currents are first class constraints.
Obstructions in gauging WZ-type
$\sigma$-models were analyzed from a Lagrangian perspective in
\cite{Hull} and related to equivariant cohomology in 
\cite{Figueroa-O'Farrill:1994dj}. 
In our approach, the anomaly cancellation conditions read
$H(v_a, \cdot, \cdot) =  \md \alpha_a$ with 
$\alpha_a(v_b) + \alpha_b(v_a)=0$.

In the WZW model, the currents (\ref{eq:KM}) form the standard
one-dimensional current algebra,
$$
\begin{array}{lll}
\{ J^L_a(\sigma), J^L_b(\tau) \} & = & 
f_{ab}^c J^L_c(\sigma) \delta(\sigma-\tau) + \frac{k}{2\pi} \delta_{ab}
\delta'(\sigma - \tau) , \\
\{ J^R_a(\sigma), J^R_b(\tau) \} & = & 
f_{ab}^c J^R_c(\sigma) \delta(\sigma-\tau) - \frac{k}{2\pi} 
\delta_{ab} \delta'(\sigma - \tau) , \\
\{ J^L_a(\sigma), J^R_b(\tau) \} & = & 0 .
\end{array}
$$
By Fourier decomposition one obtains the more familiar
form of the Kac-Moody algebra.
The combination  $J_a=J^R_a - J^L_a$ is anomaly free and can be
gauged out. In fact, this is exactly the constraint of the
gauged WZW model (\ref{eq:WZW}), (\ref{eq:G/G}),
$$
J= gpg^{-1} - p + \frac{k}{4\pi} \left(
\partial g g^{-1} + g^{-1} \partial g \right)\, .
$$
The corresponding Dirac structure is formed by pairs
$$
v=x^R-x^L \, \, , \, \, \alpha=\frac{k}{4\pi} {\rm Tr} \, x(dgg^{-1}+g^{-1}dg),
$$
where $x$ is an element of the Lie algebra and $x^L$ and $x^R$ are 
the corresponding left- and right-invariant vector fields on the group $G$.

The other example we want to discuss are the constraints
(\ref{eq:const}) of Poisson $\sigma$-models and
WZ-Poisson $\sigma$-models. These models are topological if all of the
constraints are of the first class. Consider the actions (\ref{eq:PSM}) and (\ref{eq:HPSM})
with no restriction on the background field $\Pi^{ij}$. At the Hamiltonian
level one can even relax the condition that $\Pi^{ij}$ be skew-symmetric.
Then, according to the above considerations, 
the constraints (\ref{eq:const}) are of the first class iff
the pairs $(v=\Pi^{ij}\alpha_i \partial_j, \alpha=\alpha_i dX^i)$ form a 
Dirac structure.
In the case of the action (\ref{eq:PSM}) this amounts
to the tensor $\Pi^{ij}$ being skew-symmetric and Poisson. The first
condition comes from isotropy of $D$ with respect to the scalar
product (\ref{eq:scal}) and ensures vanishing of the anomaly in the
current algebra. The second condition results from the closedness of
sections of $D$ with respect to the Courant bracket (\ref{eq:Courant})
(with $H=0$) and corresponds to the closedness of current-current
Poisson brackets or commutators.

Now let us turn to the more general action (\ref{eq:HPSM}). Again we
can start with the currents (\ref{eq:const}), possibly dropping the
requirement that $\Pi^{ij}$ is antisymmetric,  and pose the question
under what conditions they can be used as first class constraints.
We see that this is equivalent to requiring that the graph of the
two-tensor $\Pi$ defines a Dirac structure. The isotropy requirement
is the same as before (since the inner product (\ref{eq:scal}) is
unchanged) and thus is satisfied iff $\Pi^{ij}$ is skew-symmetric. But
the closure condition of the Courant bracket now shows a nontrivial
effect from the contribution of $H$; as a result one finds that
the Jacobiator of $\Pi^{ij}$ does not vanish anymore, but fulfills
equation (\ref{WZPoisson}) \cite{Ctirad, 3Poisson}.  

As a possible physical realization of WZ-Poisson structures we return
to a point particle of mass $m$ in a magnetic field
$B$.\footnote{T.S.~thanks R.~Jackiw for drawing his attention
to \cite{Jackiwmonopole}, which motivated the consideration below.}
In the presence of a magnetic charge density $\rho_m$, eq.~(\ref{eq:magnetic})
defines a non-degenerate 2-form $\omega$  which is not closed,
$\md \omega= -(e/c) \, \md B \propto \rho_{m} \, \md ^3 q \neq 0$.
Let $\Pi$ be the (negative) inverse of $\omega$. It  is
no longer Poisson, but is easily seen to satisfy (\ref{WZPoisson})
with $H \propto \md B$. Letting as usual the
Hamiltonian ${\cal H}=\vec{\pi\,}^2/2m$, one obtains the
vector field $V_{\cal H}=-\Pi(\md {\cal H}, \cdot)$.
The corresponding dynamical system $(\dot{q}, \dot{\pi})=V_{\cal H}$
reproduces the equations of motion of a point particle under the
influence of the Lorentz force generated by $B$.

\subsection{Generalized complex structures}
The labels of the current algebra $(\alpha, v)$ can be 
extended to complex valued 1-forms and vector fields on $M$.
Both the scalar product (\ref{eq:scal}) and the Courant
bracket (\ref{eq:Courant}) extend in a natural way.
The form of equation (\ref{eq:main}) remains the same as well. 
This simple extension leads to the notion of {\em generalized
complex structures} recently introduced in \cite{Hitchin} (cf.~also
\cite{Gual}).

Let $E^\dC = T^\dC M \oplus (T^*)^\dC M$ be the complexified
Courant algebroid. A generalized complex structure is
a smooth family of operators $J_X: E^\dC_X \rightarrow E^\dC_X, X\in M$,
with the following properties. First, $J^2= -1$, so that $E^\dC$ splits
into two subbundles $E^\dC=E^\dC_+ \oplus E^\dC_-$, corresponding
to the eigenvalues $i$ and $-i$ of $J$. Second, the subbundles
$E^\dC_+$ and  $E^\dC_-$ are both Dirac structures in $E^\dC$.
In other words, a generalized complex structure is a 
splitting of $E^\dC$ into a sum of two complementary Dirac subbundles.

In particular, ordinary complex structures correspond to
splittings of the form $E^\dC_+ = T^{(1,0)} M \oplus (T^{(0,1)})^* M$ and
$E^\dC_- = T^{(0,1)} M \oplus (T^{(1,0)})^* M$. Another type of
examples is given by symplectic structures. In this case,
$E^\dC_+$ consists of pairs $(v, \omega(v, \cdot))$ and
$E^\dC_-$ of pairs $(v, -\omega(v, \cdot))$.

In terms of the current algebra, a generalized complex structure
gives a splitting of all currents into two anomaly free subalgebras,
such that the anomaly terms arise only in the Poisson brackets
of currents from two different subalgebras.

\subsection{D-branes}
In the case of open strings or worldsheets with boundaries, additional
input is necessary. We do not discuss this question on the level of 
action functionals such as (\ref{eq:S3}), but instead turn directly to the
Hamiltonian picture. Our phase space now is the cotangent bundle $T^*PM$ of
paths in $M$, with endpoints attached to D-branes $D_0, D_1 \subset
M$. So, $X^i(\sigma)$ is a map from $[0,1]$ to $M$ such that $X^i(0)
\in D_0$ and $X^i(1) \in D_1$. The canonical symplectic 2-form is
\be \label{eq:om3}
\omega = \int_{0}^1 \! \delta X^i(\s) \wedge \delta
p_{i}(\s)\, \md\s  \, . \ee
In order to define an analogue of the twisted
symplectic 2-form (\ref{eq:om2}) for  $T^*PM$, in addition to the
closed 3-form $H$ we need primitives $B^0$ and $B^1$ on the
D-branes, i.e.~a 2-form $B^0$ on $D_0$ satisfying $\md B^0 =
H|_{D_0}$, and likewise so for $D_1$. Then
\ba \label{eq:om4}
\omega = \int_{0}^1 \! \delta X^i(\s) \wedge \delta
p_{i}(\s)\, \md\s  \nonumber + \2 \int_{0}^1  \!
H_{ijk}(X(\s)) \, \partial X^i
(\s) \, \delta X^j(\s) \wedge 
\delta X^k(\s) \, \md\s \; \\ +
\2 B^0_{ij}({X(0)}) \, \delta X^i(0) \wedge
\delta  X^j(0) -
\2 B^1_{ij}({X(1)}) \, \delta X^i(1) \wedge \delta
X^j(1)   \, 
\ea
defines a  symplectic 2-form on  $T^*PM$. Note that the
boundary contributions are needed in verifying the closedness
condition for $\omega$. 

Given a current $J_{(v, \alpha)}$ we need to decide whether the
boundary conditions imposed by D-branes $D_0, D_1$ preserve
the symmetry generated by this current. From a mathematical
point of view, this amounts to checking whether the differential
$\delta J_{(v, \alpha)}$ can be obtained by inserting some vector 
in the 2-form $\omega$. This gives two conditions.
First, the vector field $v$ should be tangent to the D-branes
$D_0$ and $D_1$. Second, the 1-form
$\alpha + B^0(v, \cdot)$ should vanish on $D_0$ while the
1-form $\alpha + B^1(v, \cdot)$ should vanish on $D_1$.

As a first example we consider the Poisson $\sigma$-model (\ref{eq:PSM}). 
Here, $H \equiv 0 \equiv B^0 \equiv B^1$. The constraints are again of 
the form (\ref{eq:J}) with the condition that everywhere 
 $(\a,v) = (\a, \Pi(\a ,\cdot))$ and that on the boundary $v$ is tangent 
to the respective D-brane for any $\alpha$ that vanishes upon restriction
to it, i.e.~for any $\a$ in the conormal bundle to the brane. Describing the 
respective D-brane (locally) as the level zero set of some functions $f^I$, 
where $I = 1, \ldots, \dim M-\dim D$, the set of these $\a$'s is spanned 
by $\md f^I$. The condition to be satisfied is then that 
 $\Pi(\md f^I ,\cdot)) 
\equiv \{ f^I, \cdot \}$ needs to be parallel to the surfaces $f^I =0$. 
This is recognized as the 
first class property of such surfaces. So, in agreement with 
\cite{CFbrane} we find that 
admissible  D-branes  of maximal symmetry in the Poisson $\s$-model should 
be first class or coisotropic submanifolds of the Poisson manifold
$M$. But also other D-branes are conceivable, cf.~\cite{Falcetobrane},
restricting permitted $\alpha$'s to a subset of elements of the
conormal bundle of the brane (such that $v=\Pi(\alpha,\cdot)$ is still
in its tangent bundle); they are thus recognized as
branes of less symmetry. 

As a slightly more complicated  example, we consider the 
WZW model with D-branes $D_0$ and $D_1$
two conjugacy classes in $G$. Then, the symmetries generated
by left- and right-moving currents $J_a^L$ and $J_a^R$ are broken
since the left- and right-invariant vector fields are not
tangent to conjugacy classes. But the combination
$J_a=J_a^L-J_a^R$ corresponds to $v=e^L_a-e^R_a$ which is tangent
to $D_0$ and $D_1$. If we choose $B$ such that
$$
B(x^L-x^R, \cdot)=\frac{k}{4\pi} {\rm Tr}\, x(dgg^{-1}+g^{-1}dg),
$$
the second condition will be fulfilled as well, and the symmetry
generated by $J_a$'s will  be preserved by the D-branes.

\section{Outlook}
In this paper we gave a natural derivation of
the Courant bracket in terms of a new type of
current algebras. Moreover, Dirac structures correspond
to anomaly free subalgebras of this current algebra,
and generalized complex structures give rise to a splitting
of our current algebra into pairs of anomaly free subalgebras.
In fact, all axioms of the Courant bracket (or, better, the underlying 
Courant algebroid, provided one permits also degenerate inner
products) can be shown to be equivalent
to the properties satisfied by a 
current algebra of the kind introduced in this paper. 


More complicated geometric structures can be induced
by studying the current algebra including  higher order
derivatives. For example, the Poisson brackets will
close for currents of the form,
$$
J_\psi(\sigma) = v^i p_i(\sigma) +
\alpha_i \partial X^i(\sigma) + \beta_i \partial^2 X^i(\sigma) +
\gamma_{ij} \partial X^i \partial X^j ,
$$
where $\psi=(v, \alpha, \beta, \gamma)$. The Poisson brackets
take the form,
$$
\{ J_\psi(\sigma), J_\phi(\tau) \}=
- J_{[\psi, \phi]}(\sigma) \delta(\sigma -\tau) +
J_{(\psi, \phi)}(\sigma) \delta'(\sigma - \tau) +
J_{\langle \psi, \phi \rangle}(\sigma) \delta''(\sigma - \tau) ).
$$
Here one gets three different brackets, $[\psi, \phi]$ is an
extension of the Courant bracket, $(\psi, \phi)$ is an extension
of the Courant scalar product, and $\langle \psi, \phi \rangle$ is a new
skew-symmetric scalar product. The geometric meaning of this (and higher)
structures is not yet explored.

Recently, the Courant bracket attracted a lot of attention
in connection with generalized complex geometry and
supersymmetric $\sigma$-models (cf.~{\em e.g.}~\cite{GCSSusy}). 
It is natural to expect that our current algebra admits 
supersymmetric extensions which can be useful in this context.
In the case of ordinary current algebras such extensions 
have been studied in \cite{Getzler:1993py}.

In Section 2 we provided a list of $\s$-models where examples for the 
currents (\ref{eq:J}) arise as constraints or symmetry generators. 
One may address the quest for further covariant two-dimensional models where 
such currents arise in this way. In  \cite{KSS} such a model is
provided for any maximally istotropic subbundle $D$ of 
$E=T^*M \oplus TM$.  If $D$ is a Dirac structure, one obtains 
a topological model 
 generalizing the Poisson $\s$-model and the G/G WZW model, a
 Dirac $\s$-model.

\section*{Acknowledgement}
We are grateful to P.~Bressler, A.~Cattaneo, M.~Gr\"utzmann, A.~Kotov,
U.~Lindstr\"om and V. Roubtsov for useful discussions and remarks. 
Research of A.A. was supported in part by the Swiss National Science 
Foundation. Part of this work was done during our stay at 
the Erwin Schr\"odinger Institute for Mathematical Physics.

\end{document}